\documentclass[superscriptaddress,aps,twocolumn,prl,showpacs,10pt,lengthcheck]{revtex4}

\usepackage{amsmath,amssymb}
\usepackage[latin1]{inputenc}
\usepackage[english]{babel}
\usepackage{graphicx}
\usepackage{xcolor}
\usepackage[pdftex]{hyperref}

\renewcommand{\epsilon}{\varepsilon}
\providecommand{\url}[1]{\texttt{#1}}

\begin{document}
\author{J. Tailleur}
\author{M.R. Evans}

\affiliation{SUPA, School of Physics and Astronomy, University of Edinburgh,
  Mayfield Road, Edinburgh EH9 3JZ, Scotland}

\author{Y. Kafri}
\affiliation{Department of Physics, Technion, Haifa, 32000, Israel}

\begin{abstract}
  The extraction of membrane tubes by molecular motors is known to
  play an important role for the transport properties of eukaryotic 
  cells. By studying a generic class of models for the tube
  extraction, we discover a rich phase diagram. In particular we show
  that the density of motors along the tube can exhibit shocks, inverse
  shocks and plateaux, depending on parameters which could in principle
  be probed experimentally. In addition the phase diagram exhibits
  interesting reentrant behavior.
\end{abstract}
\date{\today}
\title{Non-equilibrium phase transitions in tubulation by molecular motors}

\maketitle

\begin{figure}[t]
  \includegraphics[width=.8\columnwidth]  {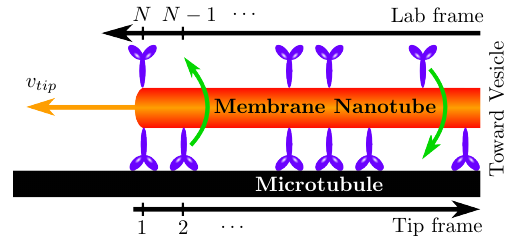}
  \caption{Illustration of tube extraction.  Molecular motors are
    attached to the membrane and can bind and unbind
    from the microtubule. Two frames of reference are
    used in the text: the `tip frame' is comoving with the tip, with
    site labels increasing toward the vesicle; the `lab frame'
    where the vesicle is stationary and the site labels increase
    towards the tip.}
  \label{fig:exp}
\end{figure}

Molecular motors play a fundamental role in intracellular
traffic~\cite{Howard2001}, being responsible for the transport of
vesicles and the extraction of membrane nanotubes~\cite{Roux2002}. The
latter phenomenon is of particular interest as it requires a
cooperative effort between many motors.  This remarkable collective
behaviour has been demonstrated in vitro only very
recently~\cite{Roux2002}, triggering much interest in the features of
the ``tubulation'', with a particular focus on the dynamics of the tip
region. Objects of study include the conditions for the formation of
tubes, their velocities, the load exerted on and by the motors, the
distribution of motors along the tube and the role of
processivity~\cite{Roux2002,Leduc2004,Campas2006,SIKSSD2008}.  In
vitro, a tube can be created when a vesicle coated with kinesins is
brought near a microtubule~\cite{Roux2002}.  It is generally
believed~\cite{Leduc2004} that two regimes are then observed depending
on the motor density: below a critical density, the motors which bind
to the microtubule are not able to extract a tube; above the critical
density tubulation occurs and the motors pull a tube out of the
vesicle, at steady velocity. During this process motors constantly
bind and unbind from the microtubule, while remaining bound to the
membrane (see figure~\ref{fig:exp}). In this regime, the density of
motors is predicted to be flat, with some structure near the tip
region~\cite{Leduc2004}. The critical density and the velocity of the
tube have been shown to be very sensitive to details of the tip
region, such as the number of motors clustered there and their
coordination~\cite{Campas2008}.

In this work, on the other hand, we focus on the regime where 
tubulation is established and study the collective behaviour of motors
in the bulk of the system. We consider a generic model of two
coupled lattices representing bound and unbound motors. By accounting
for excluded volume interactions, neglected in~\cite{Leduc2004}, we
discover a  richer phase diagram than previously expected. The
tubulation regime now divides into two different
phases with re-entrant transitions between them. Both phases
could in principle be accessible to experiment by control of the
vesicle density. As we show, the phase diagram is governed by
the bulk dynamics and the effective tip velocity; it is thus
insensitive to the precise details of the dynamics of the tip region.

\begin{figure}[t]
  \includegraphics{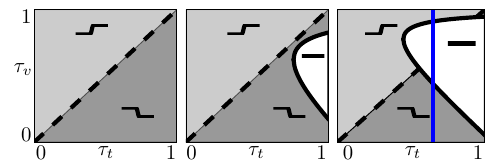}
  \caption{Possible phase diagrams depending on the value of the ratio
    $d/a$ of detachment to attachment rate. The axes represent the
    densities $(\tau_t,\tau_v)$ of bound motors in the tip and vesicle
    plateaux.  \emph{(a)} $1/4\leq d/a$. The system
    presents either shock or inverse shock profiles. \emph{(b)}
    $1/8\!<\!d/a\!<\!1/4$. The tip phase appears inside the shock
    region. \emph{(c)} $d/a\leq 1/8$. The boundary of the tip phase
    moves into the inverse shock region. In cases \emph{(b)} and
    \emph{(c)}, reentrant transitions are possible, e.g. along the
    blue line.}
  \label{fig:phasediagram}
\end{figure}

We first describe the phenomenology predicted by our study.  The motor
density profile comprises two plateaux emerging from the tip of the
tube and the vesicle, respectively. These plateaux meet in the bulk of
the system which leads to a discontinuity -- a kink -- in the density
profile. The system can be in two different phases, illustrated in
figure \ref{fig:phasediagram}: (\emph{i}) a \emph{kink} phase, in
which the tip density is either larger or smaller than that of the
vesicle, the two plateaux being connected accordingly by a shock or an
inverse shock in the bulk of the system, and (\emph{ii}) a \emph{tip}
phase where the kink travels toward {the vesicle and localizes in
  its viscinity}, thus yielding a constant density profile corresponding
to the tip density. Apart from a carefully chosen set of parameters,
the kink is never at rest, always travelling away from the tip of the
tube and either toward ({tip} phase) or away from the vesicle ({kink}
phase). In the latter case, the kink moves away from both boundaries,
which is possible because the tube is extending. The transition
between the two different phases, and also the transition from 'shock'
profile to 'inverse-shock' profiles within the kink phase, can be
triggered, for instance, by changing the value of the vesicle's
density (see the phase diagrams in figure \ref{fig:phasediagram}).
Furthermore, the phase diagram is re-entrant: by continuously
increasing the vesicle's density, one can go from the kink phase to
the tip phase and back again into the kink phase.

In non-equilibrium statistical physics, shocks play an important role
for driven lattice gas models~\cite{BE07} but the phenomenology
described here differs from previously studied biophysical traffic
problems~\cite{LW55}; reentrance is unusual and previously observed
\emph{inverse} shocks required slow particles, static defects or
special current-density relations~\cite{E95}.

\begin{figure}[t]
  \begin{center}
    \includegraphics[width=\columnwidth,viewport=59 714 310 789,clip]{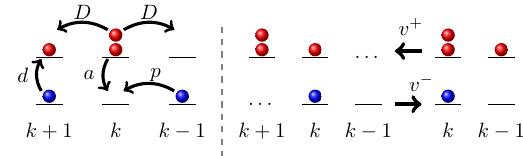}
    \caption{Coupled lattice model for the bulk dynamics. The sites
      are labelled in the lab frame. The bottom and top lattices
      represent bound and unbound motors respectively. Possible
      transitions of motors in the bulk are illustrated by arrows in
      the left panel. The right panel shows how extension and
      retraction of the tube drag the unbound motors in the bulk.}
    \label{fig:model}
  \end{center}
\end{figure}

\emph{Definition of the model}. The microscopic details of the model
are as follows (see figure \ref{fig:model}). We consider two coupled
one-dimensional lattices, for bound and unbound motors, which extend
from the vesicle to the tip of the tube. A motor bound to the
microtubule steps toward the tip of the tube at rate $p$, provided the
arrival site is empty. {Each site of the unbound motor lattice
  accounts for a whole perimeter of the tube, which in experiments
  exceeds 100 nm~\cite{Roux2002} and can contain many motors. For sake
  of clarity we thus neglect the exclusion on this lattice and assume
  that unbound motors diffuse freely at rate $D$. Partial exclusion
  could be taken into account: for realistic values of the parameters
  (see below), it does not modify qualitatively the phase diagram and
  just obscures the algebra~\cite{LongVersion}.}  Finally, motors
attach at rate $a$ to an {empty} site and detach at rate $d$ from the
microtubule.
A complete description of the tube dynamics would also include the
details of the dynamics in the viscinities of the tip and the
vesicle. As shown numerically in~\cite{Campas2008}, these details are
important for establishing the conditions for tubulation. However, as
demonstrated below, the form of the phase diagram is insensitive to
these details and relies on the fact that the tube has a well defined
mean velocity during tubulation.  We thus posit that extension and
retraction events occur with rates $v^+$ and $v^-$,
yielding a tube velocity $v_{tip}=v^+-v^-$.  Also, since the viscosity
of the membrane is two orders of magnitude larger than that of the
buffer~\cite{Leduc2004}, the unbound motors are dragged every time the
tube extends or retracts (see figure \ref{fig:model}).  Under these
conditions our results encompass a whole class of models for the tube
dynamics including, for instance, those considered
in~\cite{Campas2008}.  We now derive the different phases within a
mean-field analysis.

\emph{Mean-Field (MF) Theory.} In the analysis that follows it will be
useful to consider two distinct frames of reference: the lab frame,
where the vesicle is stationary and the site labelling starts at the
vesicle and increases toward the tip of the tube; the tip frame, which
is co-moving with the tip of the tube, and where the site labelling
starts at the tip and increases toward the vesicle (see figure
\ref{fig:exp}). In the tip frame, the mean-field equations read
\begin{equation}
    \dot\tau_i= -J_i^b+J_{i-1}^b+K_i\quad ;\quad
    \dot \sigma_i=-J_i^u+J_{i-1}^u-K_i\;.
    \label{eqn:evolTF}
\end{equation}
Here, $\tau_i$ and $\sigma_i$ are the average occupancies of bound and
unbound motors at site $i$. The current of bound motors moving between
sites $i+1$ and $i$ is given by $J_i^b= -p (1-\tau_i)\tau_{i+1} + v^+
\tau_{i}-v^-\tau_{i+1}$ whereas the current of unbound motors reads
$J_i^u= -D[\sigma_{i+1}-\sigma_i]$. Both currents are defined to be
positive in the direction of increasing $i$, i.e. when they transport
motors \emph{away} from the tip. Also, in the tip frame, extension and
retraction of the tube affects the bound motors throughout the
lattice, whence the contribution of $v^+$ and $v^-$ to $J_i^b$. Last,
$K_i=a \sigma_i (1-\tau_i)-d \tau_i$ is the flux of motors between the
two lattices.  The counterpart of these equations in the lab frame
($i$ increasing toward the tip) is easily obtained:
\begin{equation}
  \dot\tau_i= -j_{i}^b+j_{i-1}^b+K_i\quad;\quad
  \sigma_i=-j_{i}^u+j_{i-1}^u-K_i\;,
  \label{eqn:evolLF}
\end{equation}
where $j_i^u=-D(\sigma_{i+1}-\sigma_{i})+v^+ \sigma_{i}-v^-
\sigma_{i+1}$ and $j_i^b=p \tau_{i} (1-\tau_{i+1})$ are the unbound
and bound currents. Here also, currents are positive in the direction
of increasing $i$. 

As we now show, the velocity of the tip of the tube selects plateau
densities of bound and unbound motors, which we call $\tau_t$ and
$\sigma_t$ respectively. At the other end, the density of motors on
the vesicle selects in general different plateau densities which we
call $\tau_v$ and $\sigma_v$.
To derive the steady-state plateau
densities, we assume constant $\tau_{t,v}$ and $\sigma_{t,v}$ in
either \eqref{eqn:evolTF} or \eqref{eqn:evolLF}. This yields a zero
flux between the two lattices $K_i=0$, implying for any pair of
plateau densities $\tau$ and $\sigma$
\begin{equation}
  \label{eqn:Kzero}
  \sigma=d \tau/ [a (1-\tau)]\;.
\end{equation}
Let us first consider the tip plateau values $\tau_t$ and $\sigma_t$
using Eq.~\eqref{eqn:evolTF}. Adding upper and lower lattice
contributions yields a conservation equation for the total flux
\begin{equation}
  \label{eqn:currentzero}
  \begin{aligned}
F_t\equiv J_i^b+J_i^u=0 \;.
  \end{aligned}
\end{equation}
The total flux  $F_t$ flowing through the tip plateau in the tip
frame has to equal zero as nothing can move to the left of site
1. Using the explicit expressions of $J_i^b$ and $J_i^u$
in~\eqref{eqn:currentzero} and relation~\eqref{eqn:Kzero} yields
\begin{equation}
  \tau_t=1-\frac{v_{tip}}{p}\quad;\quad
  \sigma_t=\frac{d}{a}\left(\frac{p}{v_{tip}}-1\right)\;.
  \label{taup}
\end{equation}
{Note that we do not specify equation (or dynamics) near the tip
region. Solving such equations will give (generally complicated)
relations between the rates at tip region and values of $v_{tip}$,
$\tau_t$ and $\sigma_t$ while leaving Eq. \ref{taup} unmodified. They
will therefore not influence our results.}

The vesicle plateau, however, is determined by the density of motors
on the vesicle and the details of the nearby dynamics. While such
equations can be solved for specific models, to analyze the phase
diagram it is enough to know that $\tau_v$ can take any value between
$0$ and $1$.

Tip and vesicle plateau densities are typically different which
suggests the possibility of a kink phase with shock and inverse shock
profiles when $\tau_t \!>\! \tau_v$ and $\tau_t \!<\! \tau_v$
respectively. Generally, the kink is not at rest and this phase
disappears if it propagates to either end of the system. To analyse
this, we consider the kink velocity in the tip frame, $v^k_t$, and in
the lab frame, $v^k_l$. Conservation of mass implies $v^k_t=
(F_t-F_v)/(\rho_t-\rho_v)$, where $F_{t,v}$ and $\rho_{t,v}=\tau_{t,v}
+ \sigma_{t,v}$ are the \emph{total} fluxes and densities to the left
and right of the kink, in the tip frame. $F_v$ reads
    $F_v=-p(1-\tau_v)\tau_v + v_{tip} \tau_v$.
Using \eqref{eqn:Kzero} and \eqref{taup} to eliminate $\sigma_{t,v}$ and
$v_{tip}$, we obtain $v_t^{k}$ in terms of $\tau_{t,v}$:
\begin{equation}
\label{eqn:vdt}
  v_t^{k}=\frac{p \tau_v (1-\tau_t)(1-\tau_v)}{(1-\tau_t)(1-\tau_v)+d/a}\;.
\end{equation}
In the lab frame, the kink velocity is $v_l^{k}=v_{tip}-v_t^{k}$:
\begin{equation}
\label{eqn:vdl}
  v_l^{k}=p(1-\tau_t)-\frac{p \tau_v (1-\tau_t)(1-\tau_v)}{(1-\tau_t)(1-\tau_v)+d/a}\;.
\end{equation}

Since $\tau_{t,v}$ are smaller than 1, $v_t^k$ is necessarily
positive, i.e. the kink always propagates away from the
tip. Transposed in the lab frame, this means that the kink never
catches up with the tip. However, $v_l^k$ can be negative, i.e. the
kink may not propagate away from the vesicle. The tip phase indeed
occurs when the kink is localized at the vesicle and the density then
equals that of the tip plateau, except in a boundary layer close to
the vesicle.

The tip phase thus requires $v_l^{k}<0$, which reads
  $(1-\tau_v)(1-\tau_v-\tau_t) + {d}/{a}<0$ and
 can only be satisfied if
\begin{equation}
  \label{crit}
  \tau_t  > 2 \sqrt{d/a}\;.
\end{equation}
The system is then in the tip phase for $\tau_v\!\in\! [\tau_-;\tau_+]$,
where
\begin{equation}
  \label{taump}
  \tau_\pm=1-\frac{\tau_t}2 \pm \frac{\sqrt{\tau_t^2-4\,d/a}}{2}\;.
\end{equation}

When $\tau_v$ is not in $[\tau_-;\tau_+]$ or condition \eqref{crit} is
not met, the system is in the kink phase, presenting shock when
$\tau_t\!>\!\tau_v$ or inverse shock when $\tau_t\!<\!\tau_v$. Note
that for $\tau_t\!\in\![0,1]$, one always has $\tau_-\!>\!0$ and
$\tau_+\!<\!1$. The phase diagram thus always exhibits reentrance if
$d/a\!<\!1/4$, i.e. there always exist values of $\tau_t$ for which a
continuous increase of $\tau_v$ drives the system from the kink phase
into the tip phase and back into the kink phase. We present the
various possible phase diagrams in figure~\ref{fig:phasediagram}.

\emph{Numerics}. In order to validate the theoretical predictions, we
now turn to the simulation of a concrete model within the class
considered here. The bulk dynamics has already been described (see
figure \ref{fig:model}) and we now specify dynamics in the viscinity
of the tip and the vesicle. Our interest lies in verifying the phase
diagram and to this end we choose a particularly simple model. The
vesicle is represented by reservoirs of bound and unbound motors of
densities $\tau_0$ and $\sigma_0$. For simplicity, they are chosen to
satisfy \eqref{eqn:Kzero} so that there is no flux between
them~\footnote{For generic rates, a finite size boundary layer is
formed.}.  At the other end, the tube
extends at rate $\gamma$ by one lattice site if a bound motor occupies
the site next to the tip. When this happens, the new unbound site next
to the vesicle is equilibrated with the reservoir of density
$\sigma_0$. If a bound motor does not occupy the site next to the tip
the tube retracts with rate $\mu$. At this site motors can
still attach and detach with rate $a$ and $d$ and unbound motors can
hop toward the vesicle, with rate $D$. At long time, this tip dynamics
yields average extension and retraction rates $v^+=\gamma \tau_1$ and
$v^-=\mu(1-\tau_1)$.

Models aiming to predict the threshold for tubulation and tube
velocity would require modified attachment and detachment rates at the
site closest to the tip and should account for backward stepping
there. However, as noted above, the phase diagram depends only on
the tip velocity and not on further details of the tip dynamics.

\begin{figure}[t]
  \includegraphics[width=.5\columnwidth, viewport=20 8 228 153,clip]{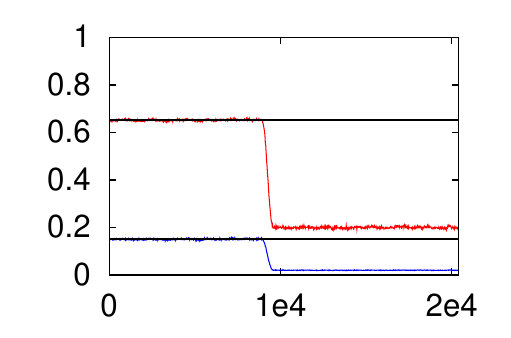}
\raisebox{.38cm}{\includegraphics[viewport=23 27 202 195, clip, angle=90,width=.45\columnwidth,totalheight=.289\columnwidth]{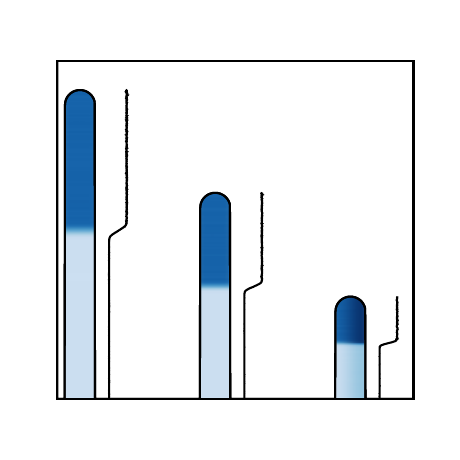}}
\caption{Shock profiles obtained from numerics ($p=1$, $d=.08$, $a=1$,
  $D=1$, $\mu=9$, $\gamma=0.45$, $\tau_0=0.2$). Data are averaged over
  a short time window ($\Delta t=1,000$) and then over 100
  simulations. \emph{Left panel:} Bound (red) and unbound (blue) motor
  densities along the tube, in the tip frame, at time
  $t=60,000$. Black lines correspond to MF predictions \eqref{taup},
  with $v_{tip}$ obtained from the simulation. \emph{Right panel:}
  Density of motors (black) at fixed time intervals ($t=4.10^4;
  8. 10^4; 12. 10^4$) in the lab frame. Tubes are color-coded, with
  light and dark blue representing low and high density. Note that the tip moves faster than the shock.}
\label{fig:profiles}
\end{figure}

We now consider the results of continuous time simulations of the
model. In figure \ref{fig:profiles}, a typical shock profile and its
dynamics are presented. Note the quantitative agreement with the
predicitions of the mean-field theory.

\begin{figure}[t]
  \includegraphics[width=.49\columnwidth, viewport=20 8 228 153,clip]{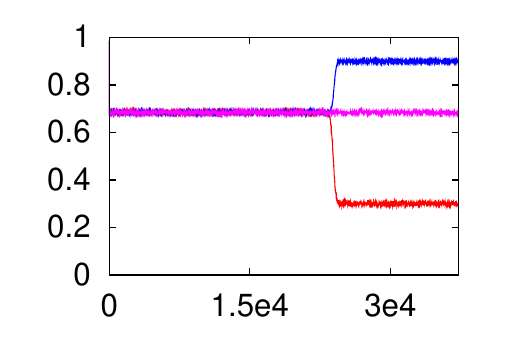}\hspace{-.25cm}
  \raisebox{-.15cm}{\includegraphics{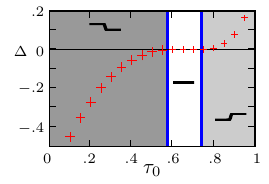}}
  \caption{\emph{Left-panel}: Different profiles are illustrated as in
    figure \ref{fig:profiles}, but with $t=120,000$, $d=0.11$ and
    $\tau_0= 0.3,\,0.6,\,0.9$. Note that the length of the lattice,
    hence the tube velocity, is independent of
    $\tau_0$. \emph{Right-panel}: Numerical value of $\Delta$
    (crosses) along the blue line of figure
    \ref{fig:phasediagram}. When $\tau_0$ goes from 0 to 1, the
    average density of bound motors is first lower than that of the
    tip plateau ($\Delta<0$, shock), then equals it ($\Delta=0$, tip
    phase), and last overcomes it ($\Delta=0$, inverse shock). The
    blue lines correspond to the predicted boundaries of the tip phase
    \eqref{taump}. There is thus a reentrant phase transition from the
    kink phase to the tip phase and back into the kink phase, whose
    boundaries are accurately predicted by the MF theory.}
  \label{fig:PhaseExemple}
  \label{Phd}
\end{figure}

We have also verified the general structure of the phase diagram and
present results for the blue line indicated in
figure~\ref{fig:phasediagram}. The different profiles observed are
presented in figure~\ref{Phd}. From previous lattice gas studies, one
would expect the inverse shock to smooth out through a rarefaction
fan. Here, on the other hand, we checked numerically that its relative
width vanishes in the large time limit; it is thus stabilized by the
interaction of the two lattices~\cite{LongVersion}. The figure also
shows that the tip velocity is independent of $\tau_v$, once
tubulation is established. A mean-field analysis~\cite{LongVersion}
suggests that this holds for generic local tip dynamics.

To quantify the transition, we define $\Delta=\sum_i \tau_i/L-
\tau_t$, where $L$ is the length of the tube. This compares the
average mass of bound motors in the system with that of a putative tip
phase. The parameter $\Delta$ is non-zero in the kink phase and zero
in the tip phase. An example of re-entrance is shown in figure
\ref{Phd}. Starting with $\tau_0$ close to 0, we see that $\Delta$ is
negative, vanishing at $\tau_0=\tau_-$ where the system enters the tip
phase. Further increase of $\tau_0$ above $\tau_0=\tau_+$ drives the
system back into the kink phase, in the inverse shock region, and
$\Delta$ becomes positive.

\emph{Conclusion} In this letter we have shown that the dynamics of
tubulation reveals a rich phenomenology, including shocks, inverse
shocks and re-entrant phase transitions. This arises from the two
competing densities set by the two ends of the tube and the dynamics
of the resulting kink determines the phase structure. This picture is
substantiated by a mean-field theory which accurately predicts the
phase diagram, as checked by our numerics.

Some experimental signatures of our theory are as follows. First, the
velocity of the tip should always exceed that of the kink. Also, once
tubulation is established, the velocity of the tip is not sensitive to
the density of motors on the surface of the vesicle. Last, in the
experiment, the ratio $d/a\simeq 0.1$ is very
small~\cite{Campas2008}. It should therefore be possible to observe
the transitions to the tip phase by varying the density of motors on
the surface of the vesicle. To explore the full phase diagram
presented in figure \ref{fig:phasediagram}, one needs to change
microscopic rates to vary $\tau_t$. Experimentally, this could be done
by changing parameters such as the membrane surface tension or the ATP
concentration. {Finally, corrections to the phase diagram due to
  partial exclusion among unbound motors is of order $d/aN_{max}$
  where $N_{max}$ is the maximal occupancy of the unbound
  lattice~\cite{LongVersion}. Here, it would be of the order of
  $0.01$.}

We thank D. Mukamel for a critical reading of the manuscript. JT
acknowledges funding from EPSRC grant EP/030173. YK thanks the Israeli
Science Foundation for support and O. Campas for discussions.

\end{document}